\documentclass[a4,11pt]{cip-submit}  
\usepackage{afterpage}

\usepackage{mathptmx}
\usepackage{hyperref}
\hypersetup{
	colorlinks=false,
	pdfborder={0 0 0}
}
\def\Mr{\uppercase}

\usepackage{float}
\usepackage{graphicx,wrapfig,tikz}
\usepackage{caption}
\usepackage{subcaption}
\usepackage{amsmath,amsxtra,amssymb,latexsym,amscd,color,bm}
\usepackage[mathscr]{eucal}
\usepackage{epstopdf}
\usepackage{array}
\usepackage{cite}

\def\vsm{\vskip0.1cm}
\def\doi#1{\href{http://dx.doi.org/#1}{doi:#1}}

\def\titles#1{\title{\large\bf\noindent #1}}
\def\authors#1{\author{\begin{flushleft}{#1}\end{flushleft}}}
\def\authord#1#2{\indent\Mr{#1}\\
	\textit{\indent#2}\vsm}

\def\email#1{\bigskip\href{mailto:#1}{\textit{E-mail:}~{#1}}\\[3mm]}

\def\received#1{\vsm\textit{\indent Received #1}}
\def\accepted#1{\vsm\textit{Accepted for publication~#1}}

\def\Keywords#1{\\[.2cm] Keywords:~{#1}.}

\def\and{$\text{\tiny AND }$}

\def\Classification#1{\\[.2cm] Classification numbers:~{#1}.} 
\begin{document}
	\Year{2015}
	\Page{1}\Endpage{9}
\titles{Gravity loop corrections to the standard model Higgs in Einstein gravity}
	\authors{	\authord{ Yugo Abe}{Department of Physics, Shimane University, Matsue 690-8504, Japan}
		\authord{ Masaatsu Horikoshi$~^\dagger$}{Department of Physics, Shinshu University, Matsumoto 390-8621, Japan}
             \authord{Takeo Inami}{Mathematical Physics Lab., Riken Nishina Center, Saitama 351-0198, Japan\\ Department of Physics, National Taiwan University, Taipei, Taiwan}
             $^\dagger$
             \email{14st307b@shinshu-u.ac.jp}
		\received{19 December 2016}
		\accepted{26 December 2016}
	}
	\maketitle
	\markboth{Y. Abe, M. Horikoshi and T. Inami}{Gravity loop corrections to the standard model Higgs in Einstein gravity}

\begin{abstract}
We study one-loop quantum gravity corrections to the standard model Higgs potential $V(\phi)$
$\grave{\rm a}$ la Coleman-Weinberg and examine the stability question of $V(\phi)$ in the energy region of Planck mass scale,
$\mu\simeq M_{\rm Pl}$ ($M_{\rm Pl}=1.22\times10^{19}{\rm GeV}$).
We calculate the gravity one-loop corrections to $V(\phi)$ in Einstein gravity by using the momentum cut-off $\Lambda$.
We have found that even small gravity corrections compete with the standard model term of $V(\phi)$ and affect the stability argument of the latter part alone.
This is because the latter part is nearly zero in the energy region of $M_{\rm Pl}$.
\Classification{04.60.Bc, 11.10.Hi, 14.80.Bn}
\end{abstract}

\bigskip

It is curious that the mass of the recently discovered Higgs boson $M_{\rm H}$ lies far outside of the mass bound derived from the one-loop radiative corrections \cite{CMPP,SZ}.
This bound arises from the stability condition on the Higgs quartic coupling $\lambda$, i.e. $\lambda(\mu)>0$.

The large two-loop corrections come into play and the renormalization group (RG) flows of $\lambda(\mu)$ change drastically. 
Some fine-tuning of the parameters, especially that of $M_{\rm t}$ $(=173.21\pm0.51\,{\rm GeV})$,
yield $\lambda(\mu)$ barely in accord with the boundary values of the stability bound extended to the scales of Planck mass $M_{\rm Pl}$ \cite{HKO,BKKS,DVEEFGA}.
This implies an interesting possibility that the standard model (SM) may hold all the way up to the Planck scale $M_{\rm Pl}$ \cite{BDGGSSS,BKPV,AFS}.
This suggestion is compatible with the so far vain results of the SUSY particle search at the LHC experiment and no experimental hints of GUT. 

It is a common belief that quantum gravity effects should manifest itself near Planck energy scales.
Possible significance of gravity effects has recently been studied in a few different approaches;
gravity loop corrections to the $\phi^6$ and $\phi^8$ terms in $V(\phi)$ in \cite{LP},
the correction to the Higgs coupling $\lambda$ in \cite{HKTY},
both in Einstein gravity and using the cut-off $\Lambda$ reguralization.
Gravity effects have also been studied by some use of string theory \cite{HKO2}.

In this letter,
we consider the graviton one-loop corrections in addition to the SM one- and two-loop corrections.
Of course the gravity one-loop corrections are small, with a coefficient of the order $(1/16\pi^{2})(\kappa^{2}\mu^{2})$ $\simeq1/10$ for $\mu\simeq M_{\rm Pl}$
(the factor $1/16\pi^{2}$ due to the one-loop integral).
However, if the standard model $V(\phi)$ becomes nearly zero at Planck scales, as advocated by many \cite{HKO,BKKS,DVEEFGA},
there is a chance that small gravity corrections may upset the SM contribution.
The gravity two-loop effects are likely to be smaller than SM one- and two-loops and graviton one-loop (to show this is an interesting question, but it is beyond this work), and will not be considered.
We are concerned about one-loop gravity effects in some energy range of $M_{\rm Pl}$.
It is conceivable that the one-loop gravity effects is small but noticeable in this energy range while two-loop effects may still be small there.
When we often say near the Planck energy $M_{\rm Pl}$, we mean $(0.1$ - $0.5)M_{\rm Pl}\leq E\leq(1.2$ - $1.5)M_{\rm Pl}$.

After taking account of the $\phi^6$ and $\phi^8$ terms in $V(\phi)$ due to gravity one-loops,
the sign of $\lambda$ alone is not enough to answer to the stability question of the Higgs potential $V(\phi)$ near the Planck energy scales, $\phi\simeq M_{\rm Pl}$.
Hence we have to study the shape of $V(\phi)$ in the region of $\phi\simeq M_{\rm Pl}$,
and we have found that $V(\phi)$ with the $\phi^6$ and $\phi^8$ terms is stable in this region.

We are aware that computing gravity loop corrections in Einstein gravity (UV non-renormalizable theory) is accompanied by some cut-off $\Lambda$ dependence.
We still think it is important to know whether gravity loop corrections to the Higgs potential may affect the results in the SM recently obtained at Planck mass scales \cite{LP, HKTY, Smolin, BM}.
On the other hand, we will see whether the $\Lambda$ dependence is rather mild so that we can say some thing useful for the study of the SM in this energy region.

We will derive the Higgs effective potential in the framework of SM coupled to Einstein's gravity theory,
the Coleman-Weinberg procedure \cite{CW}.
We begin by writing the following action,
\begin{equation}
S=\int d^{4}x \sqrt{-g}\left[-\frac{2}{\kappa^2}R+g^{\mu\nu}(\partial_{\mu}H)^{\dagger}(\partial_{\nu}H)-m^{2}H^{\dagger}H-\lambda(H^{\dagger}H)^{2}+\cdots\right],
\end{equation}
where $\kappa\equiv\sqrt{32\pi G}=\sqrt{32\pi}M_{\rm Pl}^{-1}$,
$g\equiv {\rm det}g_{\mu\nu}$,
$g_{\mu\nu}$ is the metric and $H$ is the Higgs doublet field.
The ellipsis shows the terms of gauge and fermion fields.
Expanding the Higgs doublet around the background field $\phi$ as $H^{\dagger}=1/\sqrt{2}\left(\sigma_1-i\pi_1,\phi+\sigma_2-i\pi_2\right)$
and the metric around the Minkowski background as $g_{\mu\nu}=\eta_{\mu\nu}+\kappa h_{\mu\nu}$,
we evaluate the gravity corrections to the tree level Higgs potential
\begin{equation}
V_{\rm tree}=\frac{m^{2}}{2}\phi^{2}+\frac{\lambda}{4}\phi^{4}.
\end{equation}
We take the de Donder gauge fixing term $\mathcal{L}_{\rm gf}$.
It is given in the Minkowski background by
\begin{equation}
\mathcal{L}_{\rm gf}=-\eta_{\alpha\beta}\left(\eta^{\mu e}\eta^{\nu \alpha}-\frac{1}{2}\eta^{\mu \nu}\eta^{e \alpha}\right)\left(\eta^{\rho f}\eta^{\sigma \beta}-\frac{1}{2}\eta^{\rho\sigma}\eta^{f \beta}\right)h_{\mu\nu,e}h_{\rho\sigma,f}.
\end{equation}

\begin{figure}[h!]
\includegraphics[width=60mm]{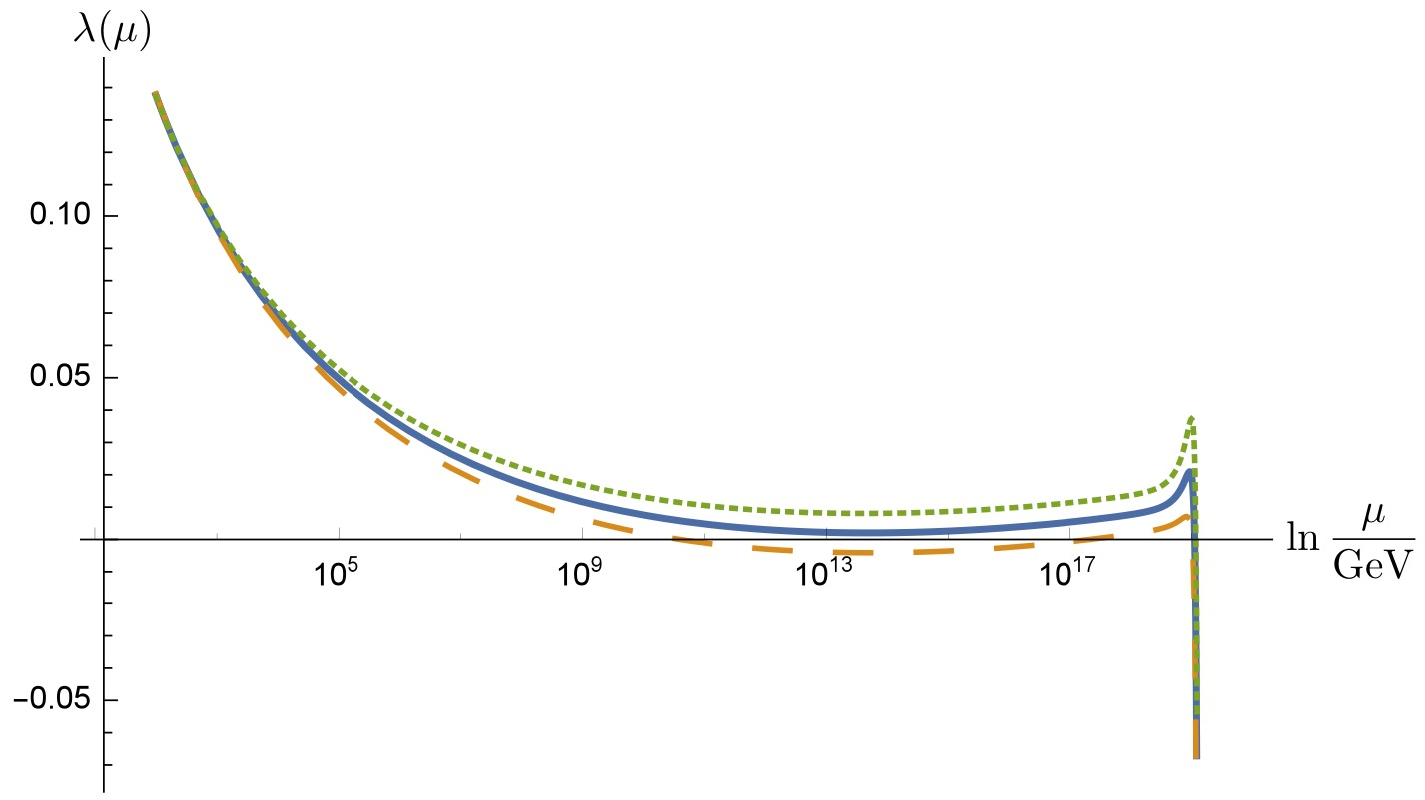}
\includegraphics[width=60mm]{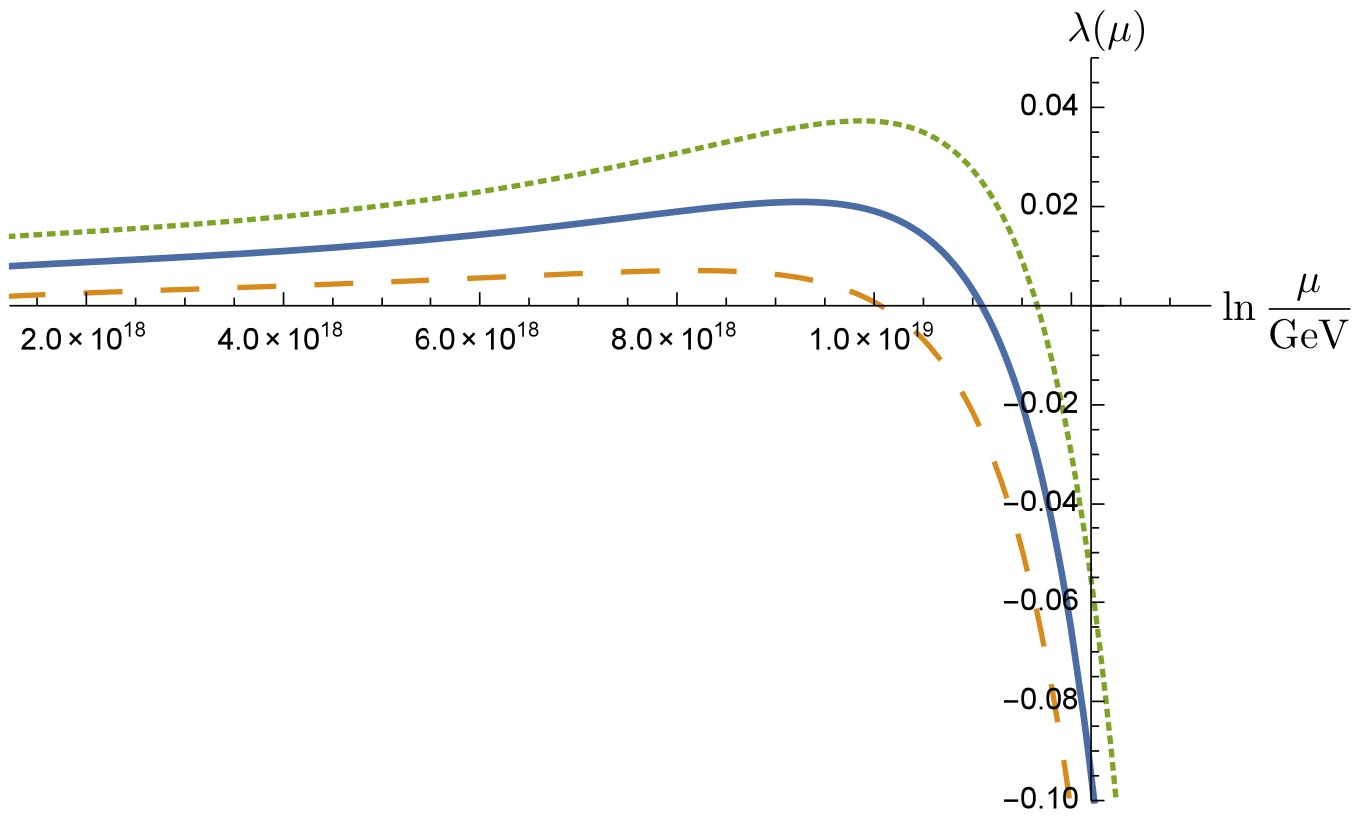}
\caption{Left: Energy dependence of $\lambda(\mu)$ for different values of $M_{{\rm t}}$, $M_{{\rm t}}=174{\rm GeV}$(dashed), $173{\rm GeV}$(solid), $172{\rm GeV}$(dotted). Right: The magnification of the Planck energy region.}
\label{fig:lambda}
\end{figure}

The gravity one-loop corrections to the potential $V_{\rm tree}$ have been obtained in the momentum cut-off method \cite{LP,BM,Smolin},
\begin{equation}
\begin{split}
\delta V_{\rm loop}=&\frac{5\kappa^{2}\Lambda^{2}}{32\pi^{2}}\left(\frac{m^{2}}{2}\phi^{2}+\frac{\lambda}{4}\phi^{4}\right)\\
&+\frac{9\kappa^{4}}{256\pi^{2}}\left(\frac{m^{2}}{2}\phi^{2}+\frac{\lambda}{4}\phi^{4}\right)^{2}\left\{\ln{\frac{\kappa^{2}\left(2m^{2}+\lambda\phi^{2}\right)\phi^{2}}{8\Lambda^{2}}}-\frac{3}{2}\right\}\\
&+\sum_{i=\pm}\frac{C_{i}^{2}}{64\pi^{2}}\left(\ln\frac{C_{i}}{\Lambda^{2}}-\frac{3}{2}\right)+\cdots,
\label{eq:potential_eff}
\end{split}
\end{equation}

where $C_{\pm}$ is
\begin{equation}
C_{\pm}=\frac{1}{2}\left[m^2_{C}-m^2_{A}\ \pm\sqrt{(m^2_{C}+m^2_{A})^2-16m^4_{B}}\right],
\end{equation}
and 
\begin{equation}
\begin{split}
m^{2}_{A}=\frac{\kappa^{2}}{8}\left(2m^{2}\phi^{2}+\lambda\phi^{4}\right),~m^{2}_{B}=\frac{\kappa}{2}\left(m^{2}\phi+\lambda\phi^{3}\right),~m^{2}_{C}=m^{2}+3\lambda\phi^{2}.
\end{split}
\end{equation}
In eq.(\ref{eq:potential_eff}),
the first and the second terms are due to the graviton one-loops,
the third term due to graviton and Higgs one-loops.
The ellipsis stands for the terms including the one- and two-loops of SM particles.
Note that the factors $\kappa^{2}$ and $\kappa^{4}$ suppress the terms at small scales of $\phi$ ($\phi\ll M_{\rm Pl}$).
Gravity corrections give rise to logarithmically divergent terms of $\phi^{6}$ and $\phi^{8}$
in addition to the $\Lambda^{2}$ and $\ln{\Lambda}$ divergences in the $\phi^{2}$ and $\phi^{4}$ terms.
The terms proportional to $\phi^{6}$ and $\phi^{8}$ are suppressed at usual energies, but they may become significant near $M_{\rm Pl}$.
The quadratic and logarithmic divergences in the $\phi^{2}$ and $\phi^{4}$ terms can be renormalized.
The effective potential $V_{\rm eff}(\phi)$ we will obtain below and will be used in this paper is the following.
\begin{equation}
\begin{split}
V_{\rm eff}(\phi)
&=\frac{m^{2}(\mu)}{2}\phi^{2}+\frac{\lambda(\mu)}{4}\phi^{4}\\
&+\frac{3}{64\pi^{2}}\left(m^{2}(\mu)+\lambda(\mu)\phi^2\right)^{2}\left(\ln\frac{m^{2}(\mu)+\lambda(\mu)\phi^{2}}{\mu^{2}}-\frac{3}{2}\right)\\
&+\frac{9\kappa^{4}}{256\pi^{2}}\left(\frac{m^{2}(\mu)}{2}\phi^{2}+\frac{\lambda(\mu)}{4}\phi^{4}\right)^{2}\left\{\ln{\frac{\kappa^{2}\left(2m^{2}(\mu)+\lambda(\mu)\phi^{2}\right)\phi^{2}}{8\Lambda^{2}}}-\frac{3}{2}\right\}\\
&+\sum_{i=\pm}\frac{C_{i}^{2}(\mu)}{64\pi^{2}}\left(\ln\frac{C_{i}(\mu)}{\Lambda^{2}}-\frac{3}{2}\right)\\
&-\frac{\kappa^{2}}{32\pi^{2}}\left(m^{4}(\mu)\phi^{2}+2\lambda(\mu)m^{2}(\mu)\phi^{4}\right)\ln\left(\frac{\Lambda^{2}}{\mu^{2}}\right)\\
&+\frac{5\kappa^{4}}{512\pi^{2}}m^{4}(\mu)\phi^{4}\ln\left(\frac{\Lambda^{2}}{\mu^{2}}\right),
\end{split}
\label{eq:full potential_eff}
\end{equation}
where $m(\mu)$ and $\lambda(\mu)$ and $C_{i}(\mu)$ are renormalized parameters.
Solving the renormalization group equation including gravity 1-loop and SM 2-loop, we have evaluated the value of these renormalized parameters.
 (see eq.(\ref{lambda's beta-function}), ref.\cite{HKO} and Fig.\ref{fig:lambda}).
Using these renormalized parameters, we can get the $V_{\rm eff}(\phi)$ including the SM one- and two-loop corrections and the gravity one-loop corrections.

\begin{figure}[h!]
\begin{center}
\includegraphics[width=120mm]{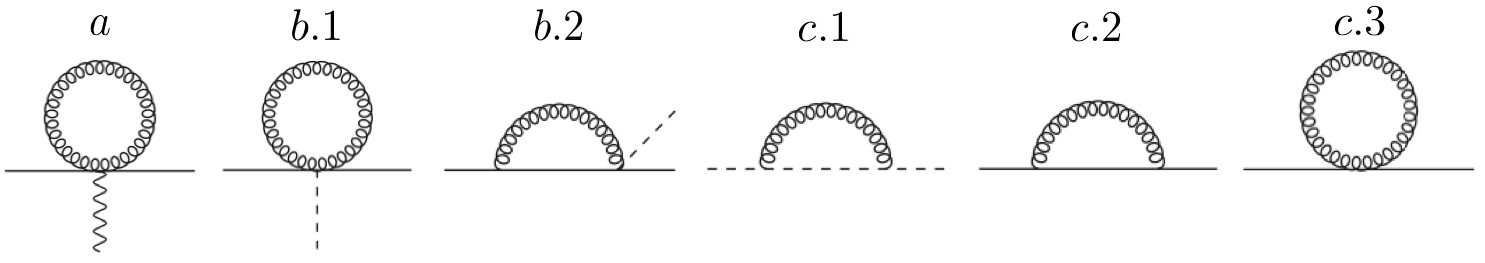}
\vspace{0mm}
\vskip-\lastskip
\caption{Gravitaton one-loop graphs for gauge couplings ($a$), Yukawa coupling ($b.1\sim b.2$), 
anomalous dimension of Higgs ($c.1$) and anomalous dimension of fermion ($c.2\sim c.3$).}
\label{fig:two}
\end{center}
\end{figure}

The $V_{\rm eff}(\phi)$ given by (\ref{eq:full potential_eff}) (here after denoted by $V(\phi)$)
receives i) the contribution from 1- and 2-loop corrections from the SM, $V_{\rm i}(\phi)$,
and ii) the contribution from the gravity 1-loop corrections, $V_{\rm ii}(\phi)$.
The term i) without gravity effects has been studied in the past \cite{HKO,DVEEFGA}.
A part of the term ii) has also been computed in \cite{LP}.
In addition the gravity 1-loop affects the coupling constants appearing in the $V_{\rm i}$, particularly $m^{2}$ and $\lambda$.
In our study the RG evaluation of these coupling constants should also take account of the gravity 1-loop corrections to the $\beta$-functions.
This will be done by considering the 1-loop graphs of Fig.\ref{fig:two},
and it yields the $\beta$-functions of eq.(\ref{anomalous dimensions}).

Admittedly there are a few insufficient aspects in our study of the loop corrected potential $V(\phi)$.
Firstly, we combine three different ways of evaluating $V(\phi)$, the SM 2-loop corrections of \cite{HKO,DVEEFGA}, the gravity 1-loop effects of  \cite{LP} and our work.
\footnote{ The computation in  \cite{HKO,DVEEFGA} is made using the dimensional regularization method, whereas we use the cut-off method. Combining the results of two different regularization methods is not totally consistent. However, the gravity 1-loop terms should not differ significantly depending on whether we use the cut-off  or dimensional method, and hence we think that our result should be reliable. Combining two different regularization methods is often used, e.g. in lattice and perturbative regularizations in QCD.}
Secondly, the initial values of the coupling constants (\ref{threshold values of cc}) are taken from the past analysis  \cite{DVEEFGA}.
It would be more appropriate to re-tune the initial values after incorporating the gravity loop effects.
Thirdly, without knowing the solid way to handle the UV non-rerormalizabe gravity loop corrections, we have taken a halfway means.
$\delta V_{\rm loop}$ of eq.(\ref{eq:potential_eff}) consists of two parts,
renormalizable $\phi^{2}$ and $\phi^{4}$ terms and non-renormalizable $\phi^{6}$, $\phi^{8}$ and non-polynomial terms.
The cutoff $\Lambda$ dependence is left in the second part as given in eq.(\ref{eq:full potential_eff}).
We have replaced the parameters $m$ and $\lambda$ in non-renormalized terms with renormalized parameters $m(\mu)$ and $\lambda(\mu)$.
Different approaches are taken for the non-renormalizable part in \cite{LP}.
Even with these few  unsatisfactory aspects of our calculation, we think our study of the Higgs potential gives a useful clue to the role of gravity corrections at Planck mass scales.

\begin{figure}[h!]
\begin{center}
\includegraphics[width=120mm]{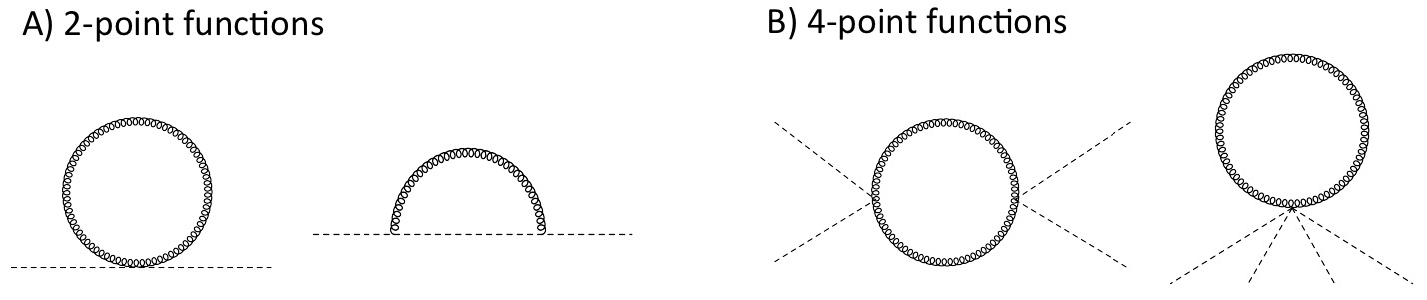}
\vspace{0mm}
\vskip-\lastskip
\caption{Graviton one-loop graphs for the Higgs two- and four-point functions.}
\label{fig:one}
\end{center}
\end{figure}

Obtaining the full effective potential (\ref{eq:full potential_eff}) amounts to finding the parameters $m(\mu)$, $\lambda(\mu)$  and $C_{i}(\mu)$ taking account of the SM two-loop corrections and gravity one-loop corrections.
This can be made by using the $\beta$-functions to these orders.
We will use the $\beta$-functions in the SM to the two-loop order 
by \cite{HKO,FJJ,CZ,BPV}.
A part of graviton loop corrections (shown in Fig.\ref{fig:one}) has been computed \cite{LP,BM,HKTY}.
\vspace{0mm}
We have further calculated gravity corrections to other coupling constants, i.e., the gauge and Yukawa couplings from graviton one-loop graphs of Fig.\ref{fig:two}.
\vspace{5mm}
We have obtained the $\beta$-functions and the anomalous dimensions due to the gravity one-loop corrections from the UV divergent terms of 
eq.(\ref{eq:potential_eff}) and these graphs.
The result is
\begin{equation}
\begin{split}
\beta_{m^2}&=\frac{5\kappa^2 m^2}{16\pi^2}\mu^{2}-\frac{\kappa^2m^4}{8\pi^2},~~
\beta_{\lambda}=\frac{5\kappa^{2}\lambda}{16\pi^2}\mu^{2}-\frac{\kappa^2\lambda m^2}{2\pi^2}-\frac{5\kappa^{4}m^{4}}{64\pi^2},\\
\beta_{g_i}&=\frac{5\kappa^2}{16\pi^2}g_i\mu^2,~~
\beta_{y_{\rm t}}=\frac{\kappa^2}{2\pi^2}y_{\rm t}\mu^2,~~
\gamma_{\phi}=-\frac{\kappa^2 m^2}{32\pi^2},~~
\gamma_{{\rm t}}=\frac{27\kappa^2}{512\pi^2}\mu^2.  
\end{split}
\label{anomalous dimensions}
\end{equation}

The energy flows of the Higgs quartic coupling $\lambda(\mu)$ and the effective potential $V(\phi)$ can be obtained by using the RG equations
with the SM matter two-loop $\beta$-functions \cite{HKO} and those due to the gravity loop corrections given in eq.(\ref{anomalous dimensions}).
We employ the threshold values of the following quantities given by Degrassi et al \cite{DVEEFGA},
\begin{equation}
\begin{split}
g_{\rm Y}(M_{{\rm t}})&=0.45187,~~g_2(M_{{\rm t}})=0.65354,\\
g_3(M_{{\rm t}})&=1.1645-0.00046\left(\frac{M_{{\rm t}}-173.15}{\rm GeV}\right),\\
y_t(M_{{\rm t}})&=0.93587+0.00557\left(\frac{M_{{\rm t}} -173.15}{\rm GeV}\right)-0.00003\left(\frac{M_{{\rm H}}-125}{\rm GeV}\right),\\
\lambda(M_{{\rm t}})&=0.12577+0.00205\left(\frac{M_{{\rm H}}-125}{\rm GeV}\right)-0.00004\left(\frac{M_{{\rm t}}-173.15}{\rm GeV}\right),
\end{split}
\label{threshold values of cc}
\end{equation}
where $g_{\rm Y}$, $g_{2}$, $g_{3}$ are the $U(1)$, $SU(2)$, $SU(3)$ gauge couplings respectively, $y_{{\rm t}}$ is the Yukawa coupling of top quark.
We adjust the value of $m^{2}(M_{{\rm t}})$ so that $V(\phi)$ gives the correct vacuum expectation value, $v=246{\rm GeV}$ at $\mu=\mathcal{O}(100{\rm GeV})$.

We investigate the following properties of $\lambda$ and $V(\phi)$.
\begin{itemize}
\item[i)]$\mu$-dependence of $\lambda(\mu)$,
\item[ii)]The shape of $V(\phi)$ near $\phi\simeq M_{\rm Pl}$,
\item[iii)]$\Lambda$ dependence of $V(\phi)$ near $\phi\simeq M_{\rm Pl}$.
\end{itemize}
In the case without gravity corrections, the RG flow of $\lambda(\mu)$ is already known\cite{HKO}.
The gravity one-loop corrections to $\lambda(\mu)$ and $V(\phi)$ is tiny at scales $\mu\ll M_{\rm Pl}$,
the RG flows of $\lambda(\mu)$ and $V(\phi)$ virtually coinciding with those of the SM only \cite{HKO} below the Planck energy scale.\\

\begin{figure}[h!]
\begin{center}
\includegraphics[width=90mm]{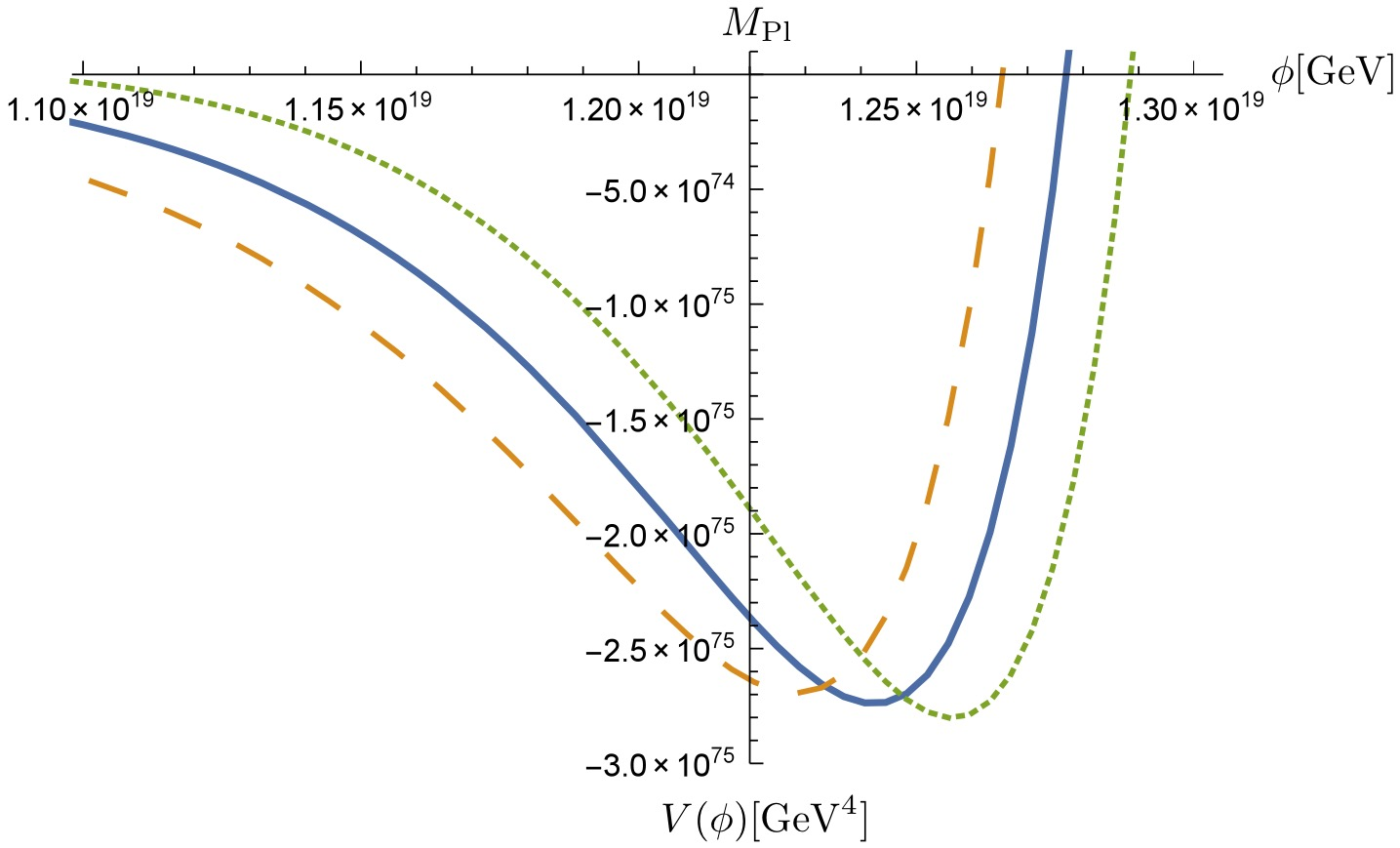}
\vspace{0mm}
\vskip-\lastskip
\caption{$V(\phi)$ near $\phi\sim M_{\rm Pl}$ for different values of $M_{{\rm t}}$, $M_{{\rm t}}=174{\rm GeV}$(dashed), $173{\rm GeV}$(solid), $172{\rm GeV}$(dotted).}
\label{fig:gcwefftop}
\end{center}
\vspace{-6mm}
\end{figure}

\leftline{i)$~\mu$-dependence of $\lambda(\mu)$}
\hspace{-3.5mm}We evaluate the running coupling $\lambda(\mu)$ by using the following equation,
\begin{equation}
\begin{split}
&\mu\frac{\partial\lambda}{\partial\mu}=\frac{5\kappa^{2}\lambda}{16\pi^2}\mu^{2}-\frac{\kappa^2\lambda m^2}{2\pi^2}-\frac{5\kappa^{4}m^{4}}{64\pi^2}\\
&+\frac{1}{16\pi^{2}}\left(24\lambda^{2}-3g_{\rm Y}^{2}\lambda-9g_{2}^{2}\lambda+\frac{3}{8}g_{\rm Y}^{4}+\frac{3}{4}g_{\rm Y}^{2}g_{2}^{2}+\frac{9}{8}g_{2}^{4}+12\lambda y_{\rm t}^{2}-6y_{\rm t}^{4}\right)\\
&+\frac{1}{\left(16\pi^{2}\right)^{2}}\left[-312\lambda^{3}+36\lambda^{2}\left(g_{\rm Y}^{2}+3g_{2}^{2}\right)-\lambda\left(\frac{629}{24}g_{\rm Y}^{4}-\frac{39}{4}g_{\rm Y}^{2}g_{2}^{2}+\frac{73}{8}g_{4}^{2}\right)\right.\\
&+\frac{305}{16}g_{2}^{6}-\frac{289}{48}g_{\rm Y}^{2}g_{2}^{4}-\frac{559}{48}g_{\rm Y}^{4}g_{2}^{2}-\frac{379}{48}g_{\rm Y}^{6}-32g_{3}^{2}y_{\rm t}^{4}-\frac{8}{3}g_{\rm Y}^{2}y_{\rm t}^{4}-\frac{9}{4}g_{2}^{4}y_{\rm t}^{2}\\
&+\lambda y_{\rm t}^{2}\left(\frac{85}{6}g_{\rm Y}^{2}+\frac{45}{2}g_{2}^{2}+80g_{3}^{2}\right)+g_{\rm Y}^{2}y_{\rm t}^{2}\left(-\frac{19}{4}g_{\rm Y}^{2}+\frac{21}{2}g_{2}^{2}\right)\\
&\left.-144\lambda^{2}y_{\rm t}^{2}-3\lambda y_{\rm t}^{4}+30y_{\rm t}^{6}\right].
\end{split}
\label{lambda's beta-function}
\end{equation}
See ref.\cite{HKO} for the details of the parts of SM.
Gravity corrections are noticeable around $\mu=\mathcal{O}(10^{18}{\rm GeV})$, with a rapid increase in $\lambda(\mu)$,
as seen from Fig.\ref{fig:lambda}.
This behavior stops at $\mu=(0.9\sim1.0)\times M_{\rm Pl}$, and $\lambda(\mu)$ starts to decrease sharply.
It becomes negative at $\mu\simeq M_{\rm Pl}$ (Fig.\ref{fig:lambda}).
\\
\leftline{ii)~The shape of $V(\phi)$ near $\phi\simeq M_{\rm Pl}$}
$V(\phi)$ near $\phi=M_{\rm Pl}$ is shown in Fig.\ref{fig:gcwefftop} (for different values $M_{\rm t}$).
We have set the condition $\mu=\phi$ to see how the potential $V(\phi)$ changes as $\phi$ varies together with  $\mu$. 
Gravitational effects begin to be noticeable at $\phi=\mathcal{O}(10^{18}{\rm GeV})$, where $\phi^6$ and $\phi^8$ terms become important.
In the region of $\phi<0.8M_{\rm Pl}$, $V(\phi)$ is positive. 
In the region of $\phi=(0.8\sim0.9)\times M_{\rm Pl}$, $V(\phi)$ begins to be negative.
At $\phi=1.1M_{\rm Pl}$, it takes a minimum.
In the region of $\phi\gtrsim 1.1M_{\rm Pl}$, $V(\phi)$ is rapidly increasing.
However, at such large values of $\phi$, higher loop effects may be important,
and one cannot say anything reliable about the hight (and the shape) of $V(\phi)$.
\vspace{0mm}

From i) and ii), we note that the graviton loop effects are essential to determine the stabilization of $V(\phi)$ in the Planck energy region.
We have already noted in Fig.\ref{fig:lambda} that $\lambda(\mu)$ becomes negative at $\mu\simeq M_{\rm Pl}$.
In the SM, $\lambda(\mu)<0$ makes the $V(\phi)$ unstable.
However, Fig.\ref{fig:gcwefftop} shows that the potential is stable thanks to $\phi^{6}$ and $\phi^{8}$ terms due to the graviton loop corrections.
Hence, the study of the potential stabilization requires the evaluation of the graviton loop effects as the energy approaches Planck mass scale.

We have so far taken a single fixed value of the cut-off, $\Lambda=1M_{\rm Pl}$ (in Fig.\ref{fig:lambda} and Fig.\ref{fig:gcwefftop}).
We know that quantum corrections in the non-renormalizable Einstein gravity is cut-off dependent and at best has only a limited meaning.
It is still useful to know whether gravity corrections are negligibly small or as large as the SM corrections for the cut-off values in question in this study, $1M_{\rm Pl}<\Lambda<3M_{\rm Pl}$.
\vspace{5mm}
\begin{figure}[h!]
\begin{center}
\includegraphics[width=90mm]{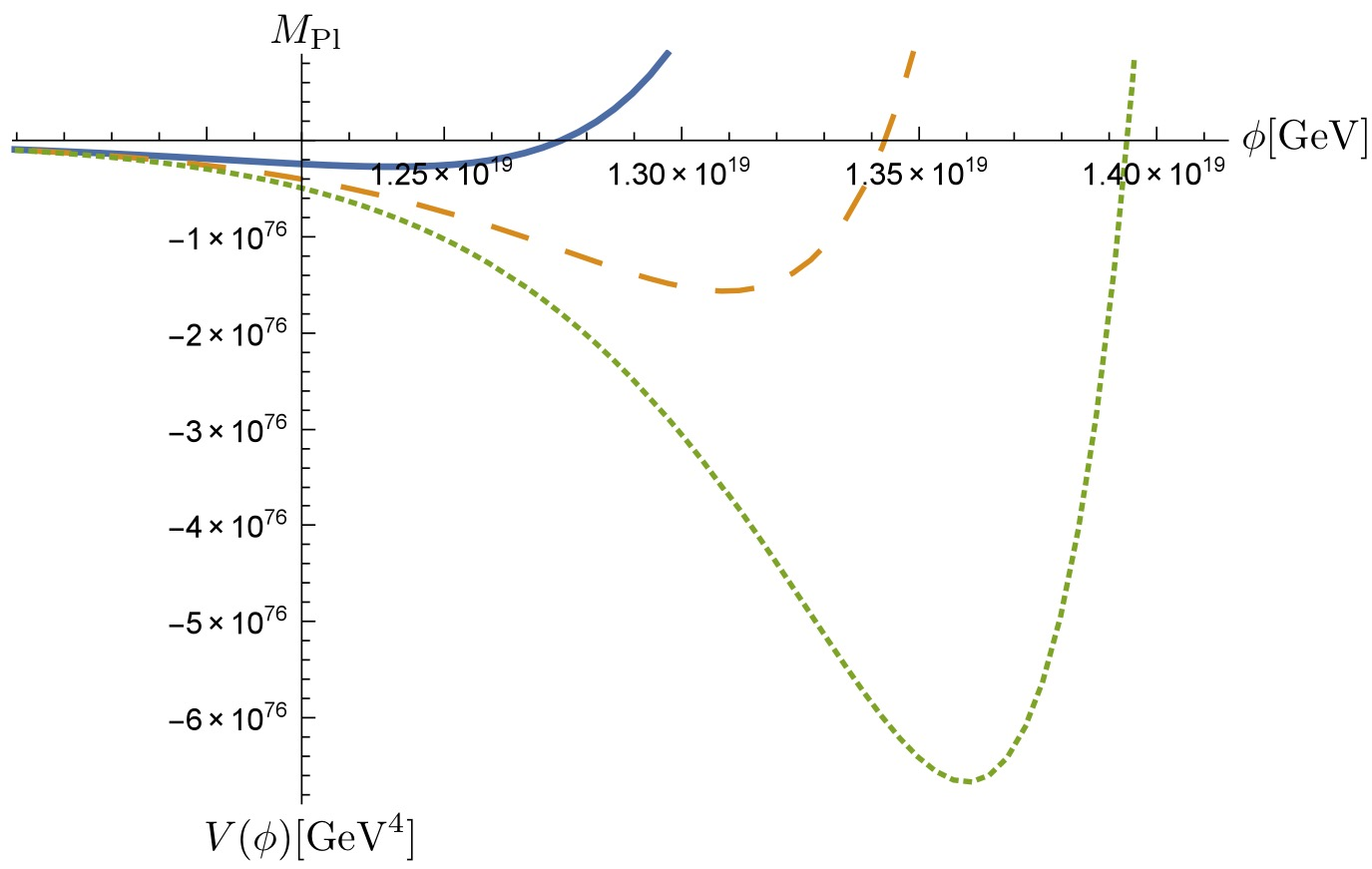}
\vspace{0mm}
\vskip-\lastskip
\caption{$V(\phi)$ near $\phi \sim M_{\rm Pl}$ for different values of $\Lambda$, $\Lambda=3M_{\rm Pl}$(dotted), $2M_{\rm Pl}$(dashed), $1M_{\rm Pl}$(solid).}
\label{fig:gcweffcut}
\end{center}
\vspace{-6mm}
\end{figure}\\
\leftline{iii)~$\Lambda$ dependence of $V(\phi)$ near $\phi\simeq M_{\rm Pl}$}
In the region of $\phi\gtrsim M_{\rm Pl}$, $\phi^6$ and $\phi^8$ terms are significant but they depend on the cut-off value $\Lambda$, as shown in Fig.\ref{fig:gcweffcut}.
$V(\phi)$ takes the minimum near $M_{\rm Pl}$, even if we change $\Lambda$.
Hence, in the graviton one-loop level, $V(\phi)$ has a minimum independent of the value of $\Lambda$.
The depth of the minimum depends strongly (at $\phi=\phi_{\rm min}$) on $\Lambda$,
but the dependence of $\phi_{\rm min}$ on $\Lambda$ is rather mild.
Hence, $V(\phi)$ takes the minimum at $\phi<\Lambda$ and $\phi_{\rm min}$ stays near $M_{\rm Pl}$
($\phi_{\rm min}$ is somewhat larger in the case of $\Lambda=1M_{\rm Pl}$, though).

\begin{figure}[h!]
\includegraphics[width=60mm]{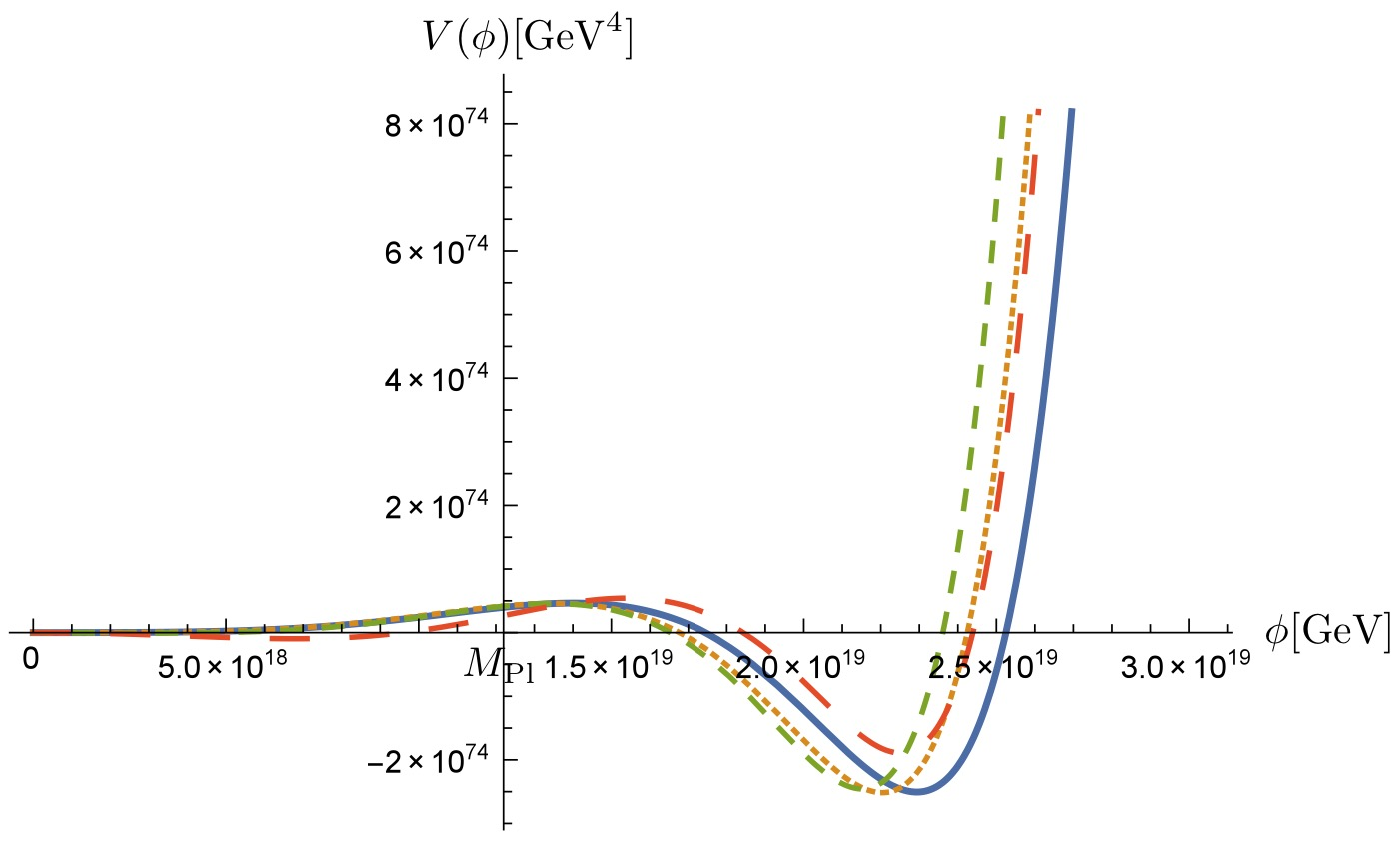}
\includegraphics[width=60mm]{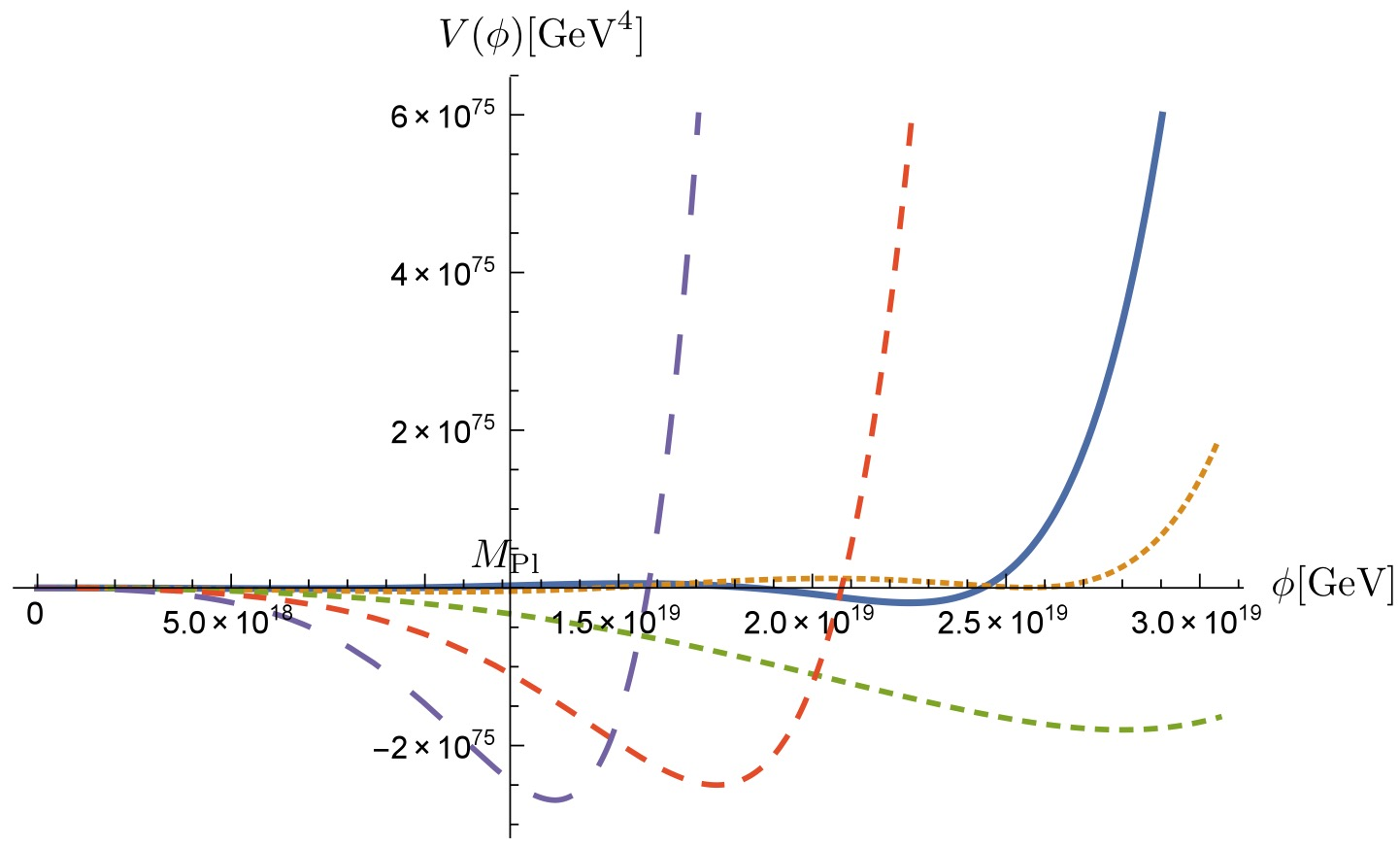}
\caption{$V(\phi, \mu)$ near $\phi\sim M_{\rm Pl}$ for different values of the energy scale $\mu$.
Left: $\mu=0.5M_{\rm Pl}$(solid), $\mu=0.6M_{\rm Pl}$(dotted), $\mu=0.7M_{\rm Pl}$(medium dashed) and $\mu=0.8M_{\rm Pl}$(long dashed).
Right: $\mu=0.8M_{\rm Pl}$(solid), $\mu=0.85M_{\rm Pl}$(dotted), $\mu=0.9M_{\rm Pl}$(small dashed), $\mu=0.95M_{\rm Pl}$(medium dashed) and $\mu=M_{\rm Pl}$(long dashed).}
\label{fig:six}
\end{figure}
Finally, we have evaluated $V(\phi)$ by setting $\mu\neq\phi$.
In Fig.\ref{fig:six}, we have compared $V(\phi, \mu=\phi)$ and $V(\phi, \mu\sim M_{\rm Pl})$.
The renormalized parameters ($m(\mu), \lambda(\mu), ...$) have $\mu$ dependence but the value of $\phi$ does not.
Presently we are not very sure how we may make our analysis more precise,
but we have compared $V(\phi, \mu=\phi)$ and $V(\phi, \mu\sim M_{\rm Pl})$ to get some measure of the stability of the potential in the Planck energy region.
This result supports our suggestion that the study of the potential stabilization requires not only the SM loop effects but also the evaluation of the graviton loop effects.

In conclusion, evaluating the quantum gravity one-loop corrections to $V(\phi)$ near $\phi\simeq M_{\rm Pl}$ in the SM coupled to Einstein's gravity theory with the momentum cut-off method,
we have found a significant difference between $V(\phi)$ with both matter and gravity loop corrections and that without gravity corrections.
The results imply that the graviton loop effects are essential to know the stabilization of $V(\phi)$ near the Planck energy scales.

In the previous work \cite{HKO2}, it is suggested that the smallness of both $\lambda$ and its $\beta$-function is consistent with the Higgs potential being flat around the string scale.
Our result agrees with this suggestion.
Actually, the gravity one-loop corrections is not significant in the region of $\phi\ll M_{\rm Pl}$.
However, in the region of $\phi\gtrsim M_{\rm Pl}$, the shape of $V(\phi)$ changes drastically by gravity loop corrections.
$V(\phi)$ with gravity corrections possesses a minimum at $\phi=\phi_{\rm min}$ somewhere $\phi\sim M_{\rm Pl}$, while $V(\phi)$ without gravity corrections increases monotonically as $\phi$ increases.
The height of the potential minimum depends on $\Lambda$ strongly,
whereas the location of $\phi_{\rm min}$ depends only weakly on $\Lambda$.
The potential minimum exists regardless of the value $\Lambda$.
In the graviton one-loop level, if the sign of $\lambda$ is negative,
the Higgs potential after including gravity corrections possesses the minimum somewhere near $\phi\simeq M_{\rm Pl}$ thanks to $\phi^{6}$ and $\phi^{8}$ terms.
A study of the Higgs vacuum metastability including $\phi^{6}$ and $\phi^{8}$ terms has recently been made without referring to gravity loop corrections \cite{BM2}.
Because the Einstein gravity is non-renormalizable,
there may be different approaches for dealing with the UV divergent quantum corrections.
For instance, Loebbert and Plefka introduce counter terms for the terms of $\phi^{6}$ and $\phi^{8}$ \cite{LP}.

It has been proposed that the standard Higgs potential with the additional $\xi R\phi^{2}$ term may play a role of cosmological inflation \cite{HKO2,BS}.
It is an interesting work to study the gravity loop corrections to Higgs field in this context \cite{Saltas}.

We should further study UV renormalizable modified gravity theories without $\Lambda$ dependence.
Indeed, it has been proposed that $R^{2}$ gravity theory is UV renormalizable \cite{Stelle,FT}.
A different approach has been taken in an early work \cite{Smolin}.
In a future work, we will consider $R^{2}$ gravity as a modest step and evaluate the gravitational Coleman-Weinberg corrections to $V(\phi)$ in a UV renormalizable gravity.
\section*{Acknowledgements}
It is our great pleasure to thank C.\,S.\,Lim and C.\,S.\,Chu for useful discussions and encouragement.
M.\,H. benefited much from his visits to National Taiwan University.
He wishes to thank P.\,M.\,Ho of National Taiwan University for hospitality and the financial support for the visits. 
We greatly appreciate valuable discussions with Y.\,Kawamura and K.\,Okuyama.
This work was done partly while we were visiting RIKEN, Tsuboi lab of Chuo University.
We wish to thank T.\,Hatsuda of RIKEN, Y.\,Tsuboi and Y.\,Sugawara of Chuo University for their kind hospitality.

\end{document}